\documentclass[10pt]{article}
\usepackage{graphicx}
\begin{document}

\vspace*{5mm}

\begin{center}
{\bfseries
\uppercase{Setup for the Nuclotron beam time structure measurements}
}

\vspace*{5mm}

A.Yu.~Isupov$^{\dag,\ddag,}$\footnote{Corresponding author.},
V.P.~Ladygin$^{\dag}$,
and S.G.~Reznikov$^{\dag}$

\vspace*{5mm}

{\small {\it $^\dag$ Joint Institute for
Nuclear Research, 141980 Dubna, Russia}} \\
{\small {\it $^\ddag$ E-mail: isupov@moonhe.jinr.ru}}\\
\end{center}

\vspace*{5mm}

\centerline{\bf Abstract}
The setups
 for precise measurements  of the time structure of Nuclotron
internal and slowly extracted beams
are described in both hardware and software aspects.
The CAMAC hardware is based on the use of the
standard
CAMAC modules developed and manufactured at JINR.
The data acquisition
system software is implemented using the {\itshape ngdp} framework under the
Unix-like operating system (OS) FreeBSD to allow the easy
network distribution of the online data. It is demonstrated that
the described setups are suitable for the continuous beam quality monitoring
during the experiments performed at Nuclotron.

\vspace*{3mm}

Keywords:
CAMAC,  data acquisition system,
 network distributed system,  graphic user interface,
 beam time structure

\vspace*{3mm}

{\textbf{PACS:~29.27.Fh, ~07.05.Hd}}



\newpage

\setcounter{page}{1}

\section*{Introduction
}
\label{spill.intro}

The experimental program at Nuclotron for heavy ion collisions
\cite{bmn_CDR,LadyISHEPP12}, few body \cite{Ladygin:2014yla} and  polarization \cite{dss1,bmn_baldin2014}
physics requires good quality of the internal and  extracted beams. 
The time structure of the beam plays a crucial role in the experiments requiring
high interaction rate.  One of the parameters reflecting the  beam time structure is 
the   coefficient $K_{\mbox{\tiny{dc}}}$ defined as
\begin{eqnarray}
\label{eq_Kdc}
K_{\mbox{\tiny{dc}}} = \frac{1}{t_2-t_1}\cdot \left[ \int^{t_2}_{t_1} ({dN/dt}) dt \right] ^2
/ \left[ \int^{t_2}_{t_1} \left( {dN/dt} \right)^2 dt \right] \ ,
\end{eqnarray}
where $dN/dt$ is the beam current, $(t_2-t_1)=T_{ext}$ is the time of the
beam exposition. 
The real experiment requires that the $K_{\mbox{\tiny{dc}}}$ coefficient should
be about 0.85---0.95 for the efficient data taking. 

Another problem of the experiment 
related with the  beam time structure 
is the correct estimation of the data acquisition dead-time. 
This is important not only for the cross section evaluation, but also
for the polarization measurements due to effect of the dead-time distortion
\cite{dtime}.  For instance, large dead-time of the data acquisition (DAQ) due to oscillating
interaction rate could significantly reduce the asymmetry values during the measurements of 
the analyzing powers \cite{PLB2012,PPNL2011}. Such effect is especially important 
for the polarization measurements requiring the absolute normalization of the beam 
intensity or interaction rate in the wide range as in the TPD experiment on
the measurements of 
the induced tensor polarization of deuteron beam travelling through matter
\cite{AzhgPEChAJa08,AzhgPEChAJa10}. 

The Nuclotron internal target station (ITS) is an efficient tool to
perform nuclear physics studies at energies from hundreds 
of MeV to several GeV per nucleon with both polarized and 
unpolarized beams \cite{oldIntTarg}. However, in this case the useful events yield 
is the time dependent function  of the beam and moving internal target interaction.  
Recently, new system of the ITS control 
and data acquisition has been developed \cite{IsupNIM13}.     
In particular,  the internal target mechanical vibrations
were studied in details using this new system \cite{IsupNIM13}. 

The Nuclotron extracted deuteron beam time structure has been studied
previously during data taking of the TPD experiment
\cite{AzhgPEChAJa08,AzhgPEChAJa10}, on the measurements of
the soft photons yield \cite{Kokoulina2014}, and by the BM@N experiment
\cite{bmn_CDR}. The CAMAC and VME \cite{vme} based DAQ systems were used
for the measurements performed at
4V \cite{AzhgPEChAJa08,AzhgPEChAJa10,Kokoulina2014} and 6V \cite{bmn_CDR}
beamlines, respectively. 
These measurements were performed in the trigger mode of the DAQ systems,
when the information 
was read for each trigger. However, such mode cannot provide correct information
on the beam time structure at high trigger rate due to dead-time distortion. 
These circumstances motivate us to organize the setup with negligible
dead-time and to use it for the Nuclotron beam time structure measurements.

The goal of the paper is to report the details of the setups for 
the Nuclotron internal and extracted beam time structure   
measurements as well as first results obtained with deuteron, $^7$Li, and
$^{12}$C beams.  The paper is organized in the following way. The
 chapter~\ref{spill.setup} contains 
the details of the hardware and software of the setups for the Nuclotron beam time structure 
investigations. The results on the extracted beam time structure are presented in 
 chapter~\ref{spill.run_extr}. The results on the  investigation of the internal beam-target interaction are
reported in
 chapter~\ref{spill.run_intr}. The conclusions are drawn in the last chapter.

\begin{figure}[htb]

\vspace*{-5mm}

\begin{center}
\includegraphics[width=0.7\textwidth]{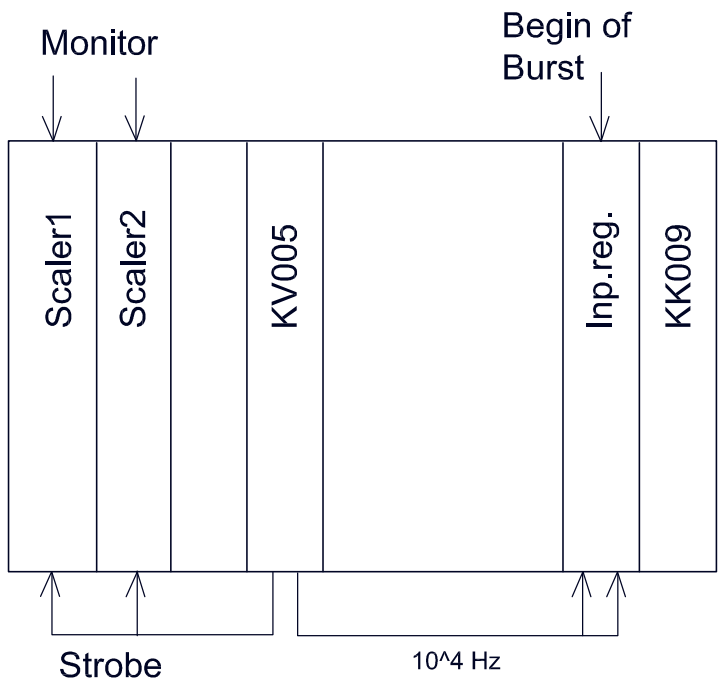} 
\end{center}

\vspace*{-7mm}

\caption{CAMAC hardware. See text for description.}
\label{spill.fig.func_scheme}
\end{figure}

\begin{figure}[htb]
\includegraphics[width=\textwidth]{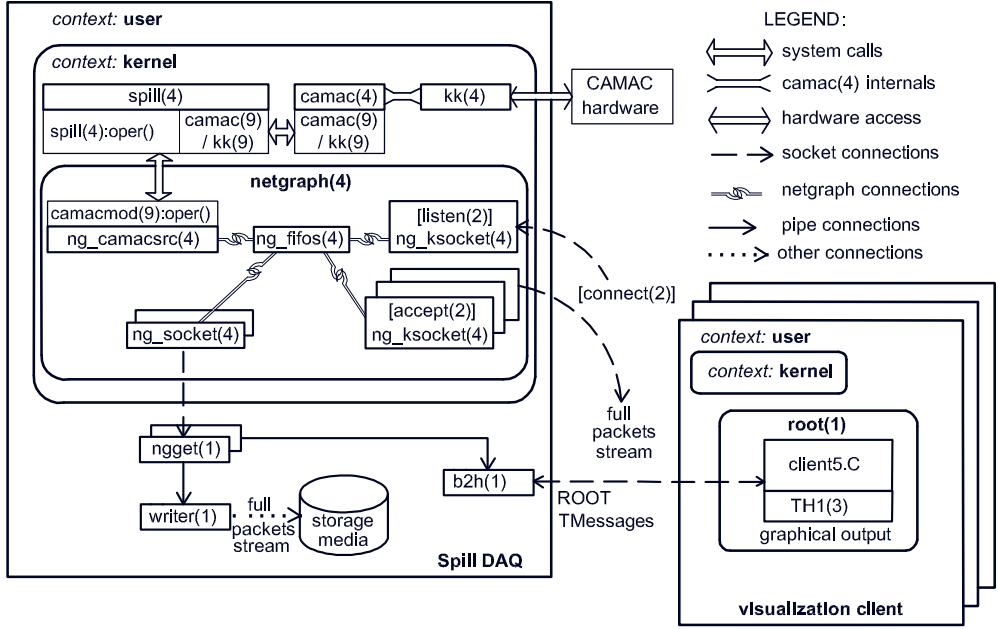} 
\caption{Overall Spill DAQ layout. See text for description.}
\label{spill.fig.schema}
\end{figure}

\begin{figure}[htb]
\includegraphics[width=\textwidth]{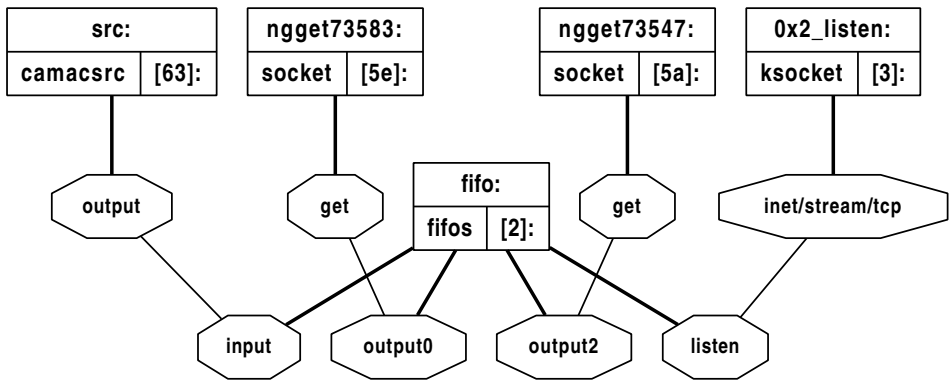}
\caption{
The Spill DAQ core is implemented by the {\itshape ngdp} graph. See text
for description.
}
\label{spill.fig.graph}
\end{figure}

\begin{table}[htb]


\caption{The user commands to control over Spill DAQ.}
\label{spill.tbl.Make}
\begin{tabular}{p{0.17\textwidth}p{0.74\textwidth}}
\hline
Command & Description \\
\hline
{\ttfamily load} & loads and configures the
  {\bfseries\itshape spill(4)} and \verb|ng_camacsrc| kernel modules,
  connects latter to \verb|ng_fifos| input \\
{\ttfamily unload} & counterpart for previous \\
{\ttfamily loadw} & loads the {\bfseries\itshape writer(1)} module
  and connects it to \verb|ng_fifos| output. \verb|RUNFILE| and \verb|DATADIR|
  variables could be defined \\
{\ttfamily unloadw} & counterpart for previous \\
{\ttfamily loadb2h} & loads the {\bfseries\itshape b2h(1)} module
  and connects it to \verb|ng_fifos| output. \verb|RUNFILE|
  variable could be defined \\
{\ttfamily unloadb2h} & counterpart for previous \\
{\ttfamily continue} & starts the handling of CAMAC interrupts \\
{\ttfamily swstart} & starts the handling of CAMAC interrupts with software
  BoB imitation by {\bfseries\itshape callout(9)} mechanism \\
{\ttfamily pause} & counterpart for two previous \\
{\ttfamily init} & initializes the CAMAC hardware \\
{\ttfamily finish} & counterpart for previous \\
{\ttfamily cleanall} & resets all histograms in the {\bfseries\itshape b2h(1)} \\
{\ttfamily saveall} & stores all {\bfseries\itshape b2h(1)} histograms into
  ROOT \verb|TFile| on hard disk (if \verb|RUNFILE| variable was
  supplied for \verb|loadb2h|) \\
{\ttfamily disconn} & disconnects all clients from {\bfseries\itshape b2h(1)} \\
{\ttfamily status} & outputs status messages from present modules through
  {\bfseries\itshape syslogd(8)} \\
{\ttfamily seelog} & starts the {\bfseries\itshape syslogd(8)} output file
  {\itshape qdpblog} viewing by {\bfseries\itshape tail(1)} \\
\hline
\end{tabular}
\end{table}

\section{Setup hardware and software}
\label{spill.setup}

\subsection{CAMAC electronics}
\label{spill.setup.hw}

Each setup contains
 input register,
two identical scalers, and pulse generator
CAMAC modules.
Their functional scheme is outlined in Fig.~\ref{spill.fig.func_scheme}.
All CAMAC units were developed and produced at JINR.
Both setups have the same modules disposition in the read-out crate.

The input register collects three inputs --- accelerator cycle begin
and two triggers --- which lead to the LAM assertion.
The input register is the only LAM source in the read-out
crate, and the KK009 crate controller \cite{KK009} is programmed to set the
interrupt request (IRQ) at each LAM occurrence. Each of this IRQ leads to the
invocation of the CAMAC interrupt handler function from the \verb|spill|
loadable kernel module (see chapter~\ref{spill.setup.sw}).
Each of two triggers corresponds to scaler module should be read-out at current
interrupt handler invocation. The two triggers should alternate strictly, so
the Spill DAQ is able to handle unproper trigger bit combinations and
trigger sequences.

The KV005 \cite{KV005} generator was used to produce the trigger signals
sequence with the base frequency of $10^4$~Hz,
which corresponds to the time slice of 100~$\mu$s.

The beam-target interaction intensity is measured by
monitor scintillation counter,
whose signal is split to fed the 0th input of both scalers.
The odd scaler counts during odd time slice and is read-out during even one,
and the even scaler --- vice versa. This scheme allows us to avoid
undercounts, only up to 1 pulse could be erroneously dropped or added in each time
slice.

Because we read from scalers the least 16 bits only at each 100~$\mu$s
time slice, the setups are able to consume counts with rate up to
$\approx 6.5 \times 10^{8}$~Hz.
This is more than enough because scalers are able to input up to
$\approx (0.7 \div 1) \times 10^8$~Hz.


\begin{figure}[htb]

\vspace*{2mm}

\includegraphics[width=0.49\textwidth]{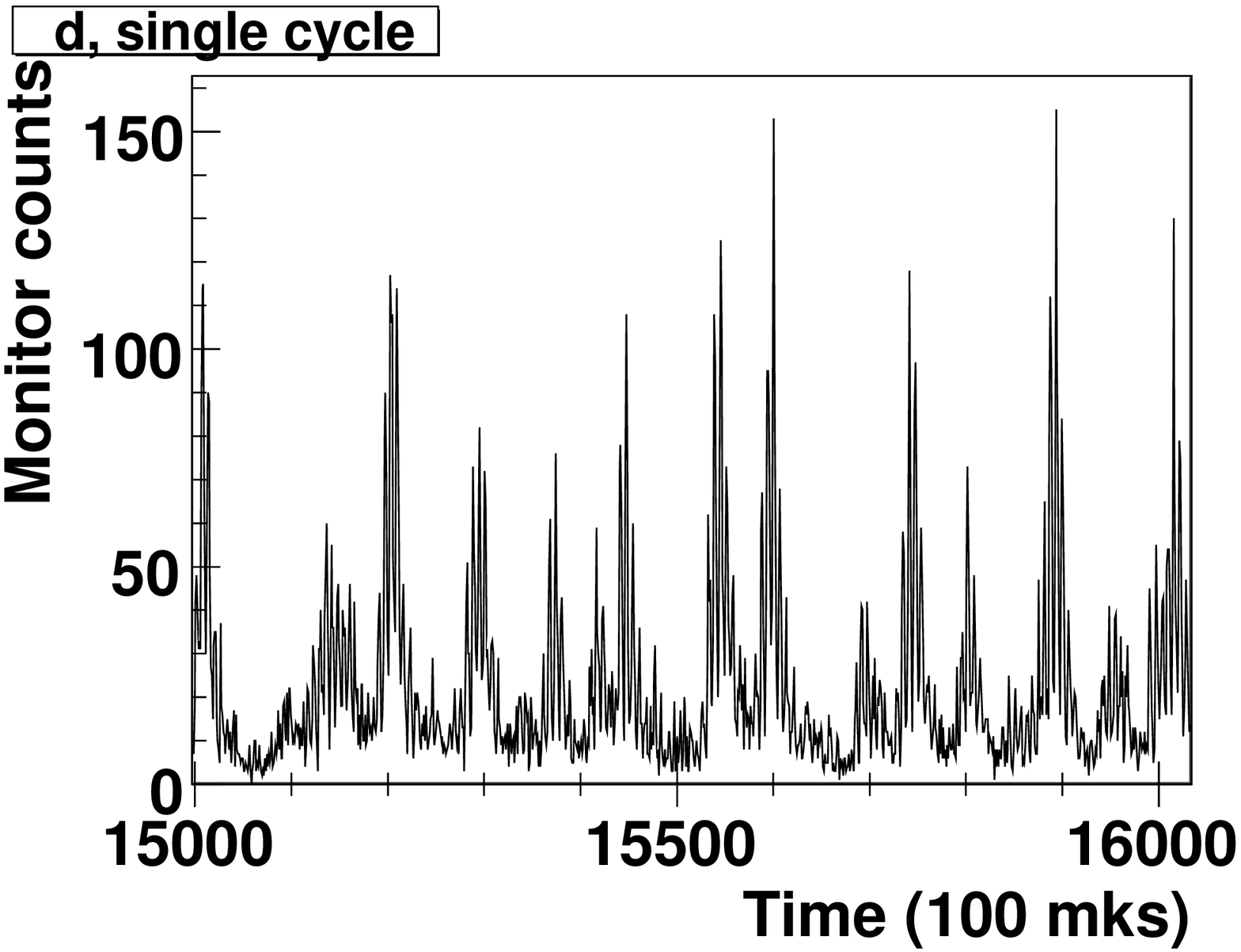}%
\includegraphics[width=0.49\textwidth]{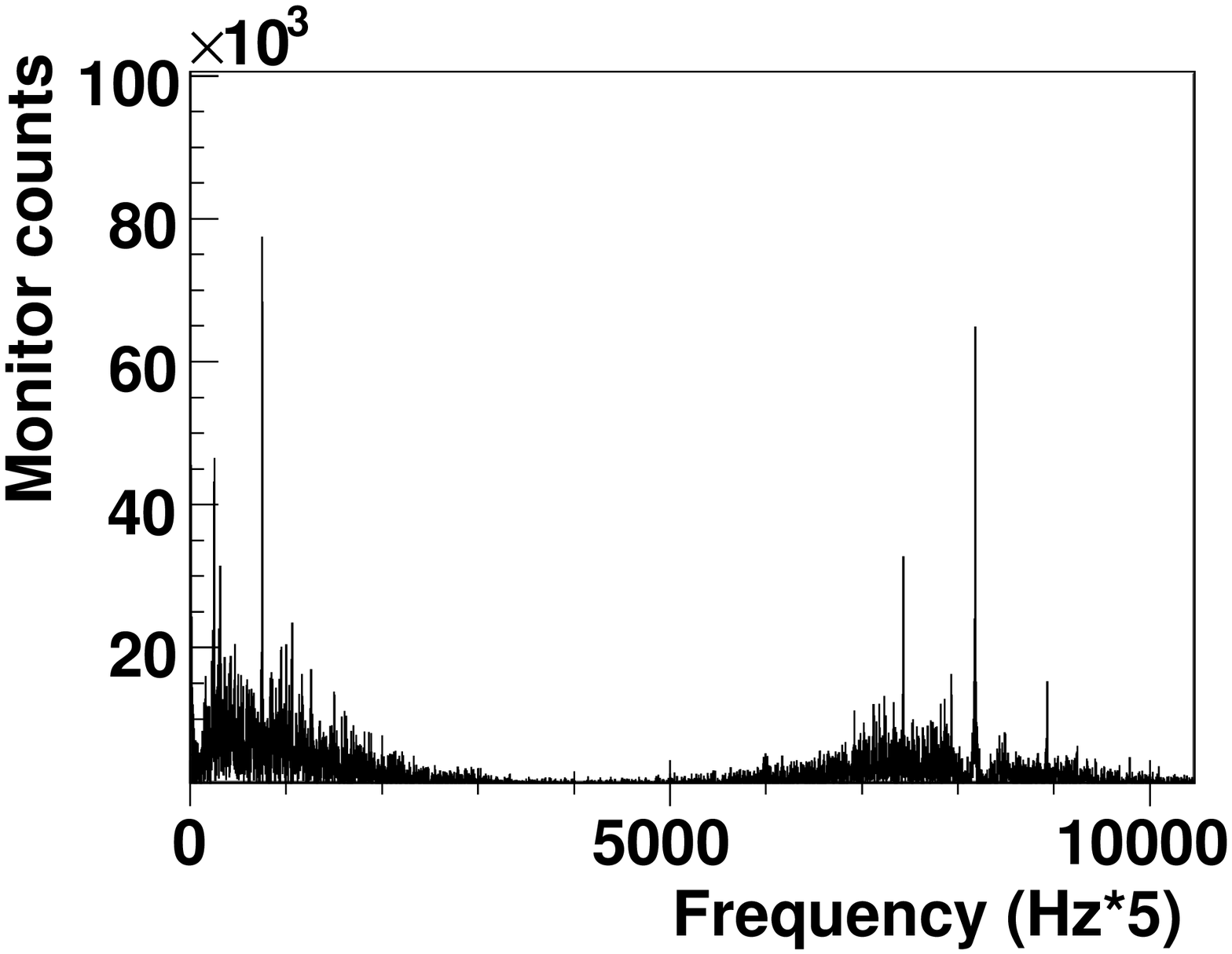}
\caption{The monitor counts in the 0.1~s window (left) and Fourier transform
of the full burst (right) during the single cycle
from the slowly extracted deuteron beam of
$T_{\mbox{\tiny{kin}}} = 3.5$~A$\cdot$GeV in F6 focus.
See text for description.}
\label{spill.fig.cnts_ext_d}
\end{figure}

\begin{figure}[htb]

\vspace*{2mm}

\includegraphics[width=0.49\textwidth]{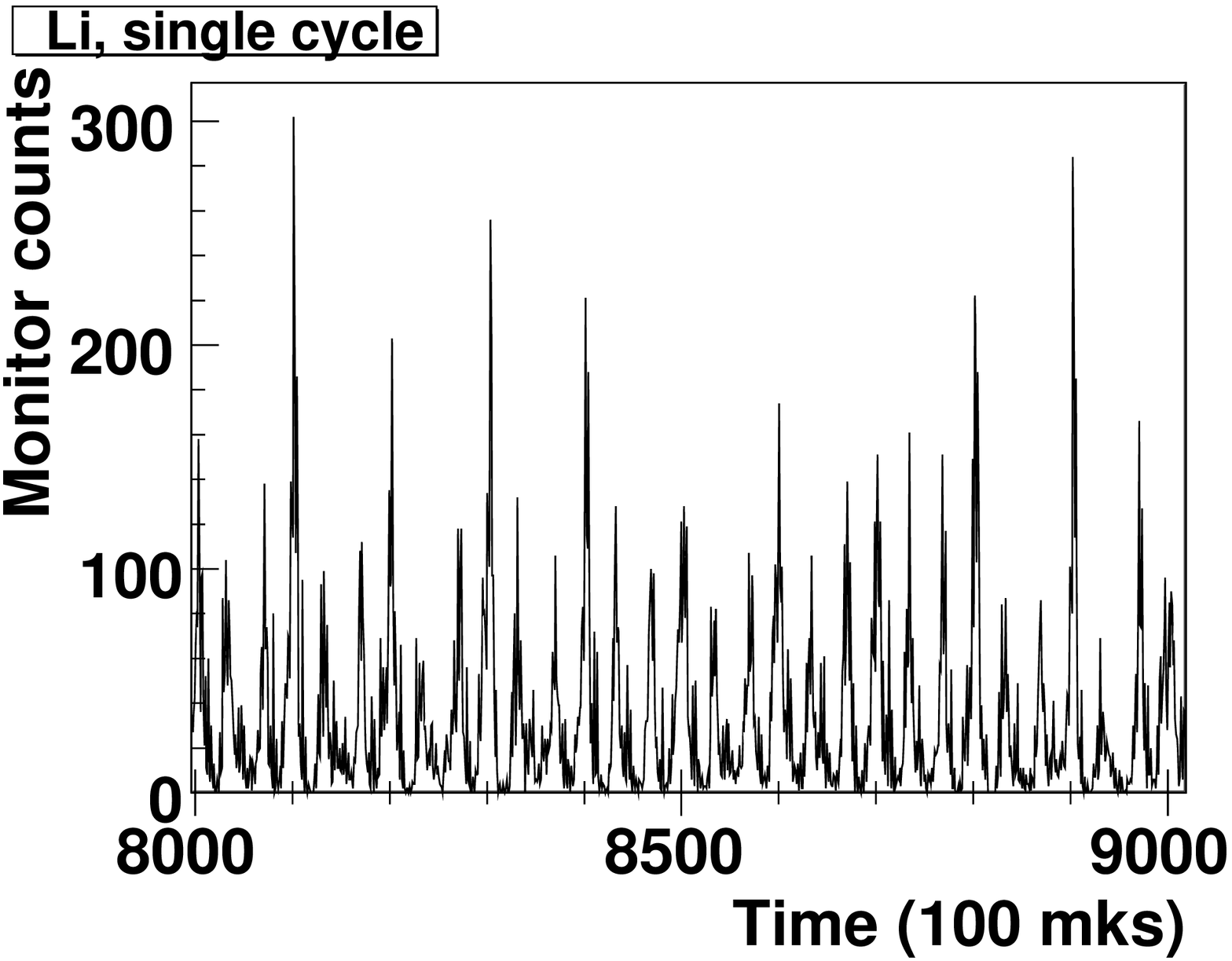}%
\includegraphics[width=0.49\textwidth]{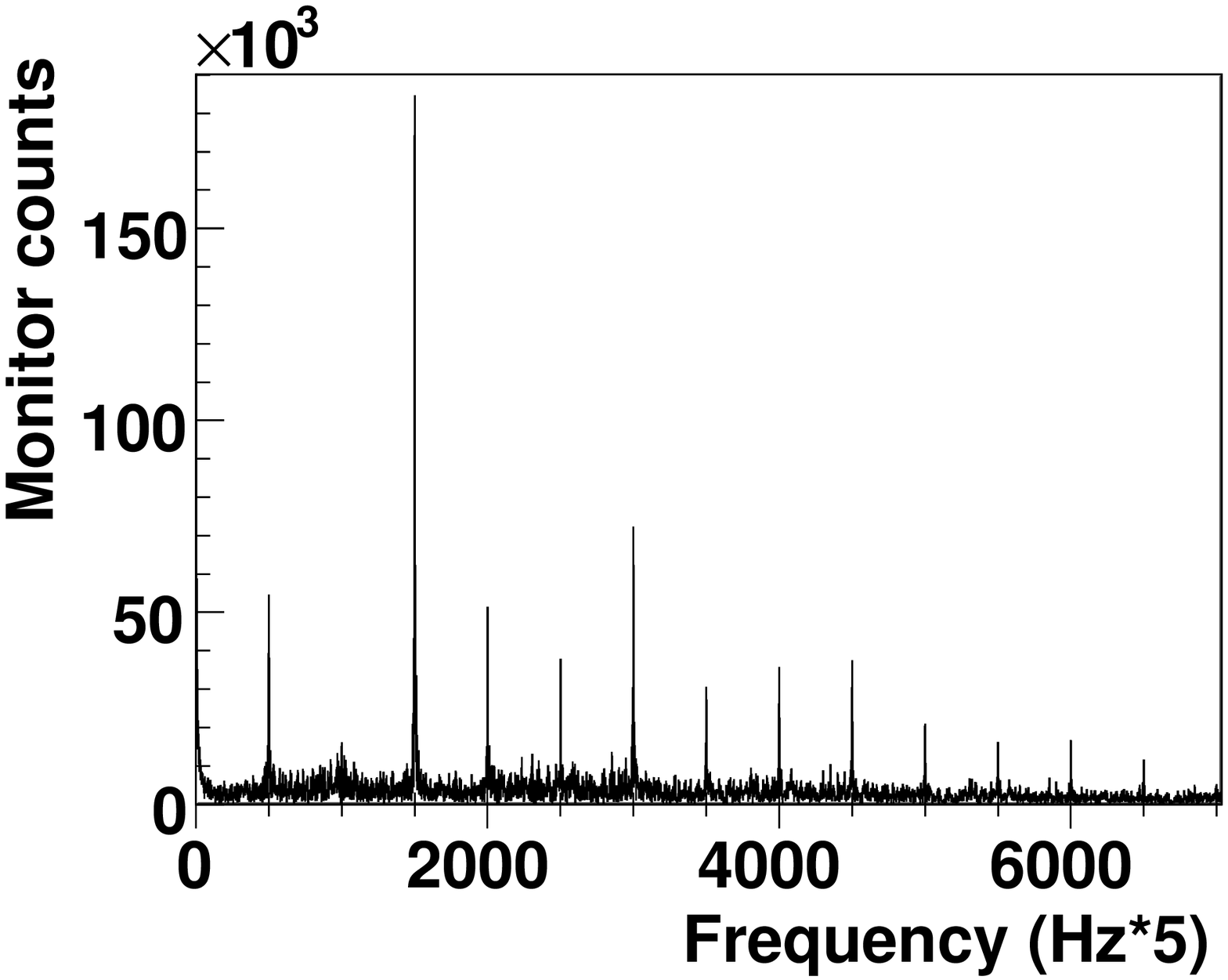}
\caption{The monitor counts in the 0.1~s window (left) and Fourier transform
of the full burst (right) during the single cycle
from the slowly extracted $^7$Li beam of
$T_{\mbox{\tiny{kin}}} = 3.5$~A$\cdot$GeV in F4 focus.
See text for description.}
\label{spill.fig.cnts_ext_Li}
\end{figure}

\begin{figure}[htb]

\vspace*{2mm}

\includegraphics[width=0.49\textwidth]{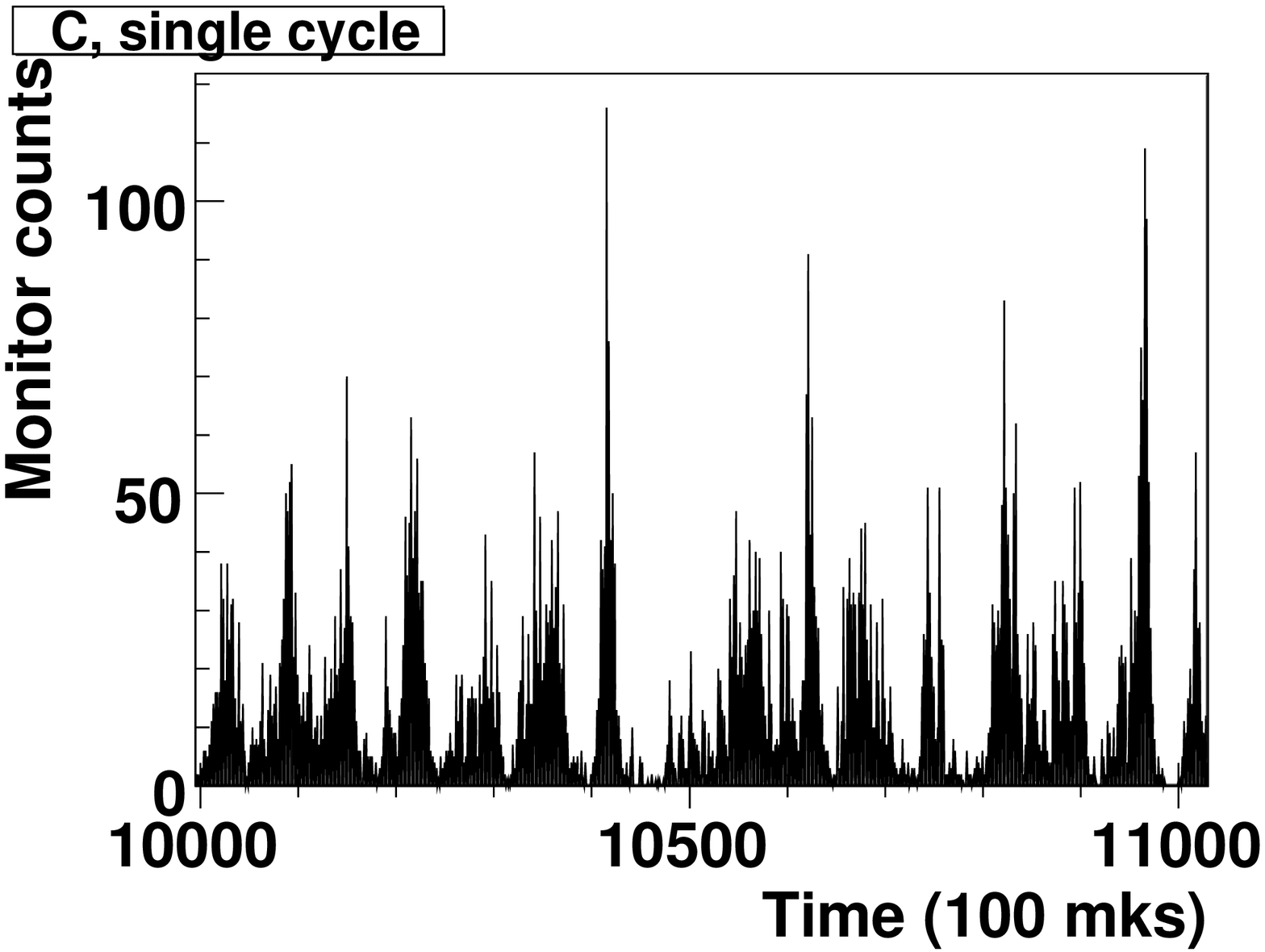}%
\includegraphics[width=0.49\textwidth]{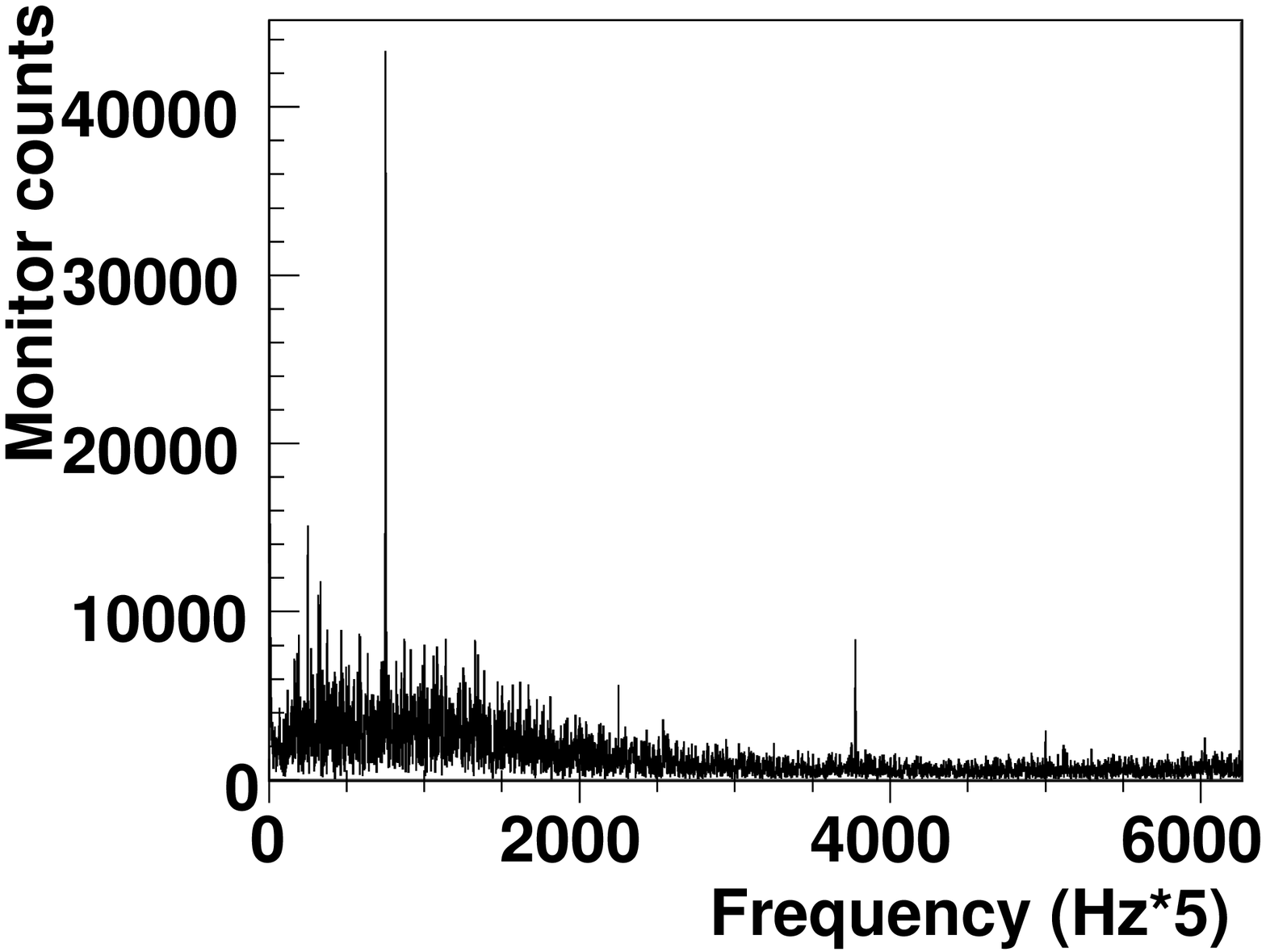}
\caption{The monitor counts in the 0.1~s window (left) and Fourier transform
of the full burst (right) during the single cycle
from the slowly extracted $^{12}$C beam of
$T_{\mbox{\tiny{kin}}} = 3.5$~A$\cdot$GeV in F6 focus.
See text for description.}
\label{spill.fig.cnts_ext_C}
\end{figure}

\begin{table}[htb]


\caption{
The time structure characteristics for the
Nuclotron slowly extracted beam.
}
\label{spill.tbl.kdc_ext} 
\begin{tabular}{rp{0.04\textwidth}%
p{0.04\textwidth}p{0.04\textwidth}%
p{0.025\textwidth}p{0.07\textwidth}p{0.07\textwidth}p{0.08\textwidth}%
p{0.09\textwidth}p{0.10\textwidth}}
\hline
Beam, $T_{\mbox{\tiny{kin}}}$ & $K_{\mbox{\tiny{dc}}}$ & $t_1$ & $t_2$
 & \multicolumn{5}{c}{\rule{0mm}{5mm}$\frac{\Delta N(\Delta f)}{0.5 N_{\mbox{\tiny{tot}}}}$~(\%)} \\
(A$\cdot$GeV) &  & (ms) & (ms)
 & \multicolumn{5}{c}{for $\Delta f$ as $f_1$..$f_2$~(Hz)} \\
\hline
\multicolumn{4}{c}{}
 & \rule{0mm}{5mm}0..1 & 100..400 & 400..800 & 800..1200 & 1200..1600 & 1600..2000 \\
\hline
d, 3.5 & 0.49 & 1000 & 2700
 & 2.4 & 21.0 & 6.48 & 5.1 & 16.4 & 11.56 \\
C, 3.5 & 0.218 & 900 & 2600
 & 1.3 & 16.0 & 7.56 & 4.63 & 6.3 & 5.1 \\
$^7$Li, 3.5 & 0.374 & 300 & 1400
 & 2.0 & 15.1 & 15.86 & 10.66 & 5.82 & 4.27 \\
\hline
\end{tabular}
\end{table}

\begin{table}[htb]


\caption{
Nuclotron internal beam--target interaction characteristics.
The time limits for the $K_{\mbox{\tiny{dc}}}$ (see equation~(\ref{eq_Kdc}))
calculations
are: $t_1 = 1000$~ms, $t_2 = 3500$~ms. The frequency ranges for the
mechanical vibrations
of the internal target ($\Delta f_{\mbox{\tiny{ITS}}}$) are
12..14~Hz for C and 13..15~Hz for CH$_2$.
}
\label{spill.tbl.kdc_int} 
\begin{tabular}{rp{0.07\textwidth}p{0.05\textwidth}%
p{0.07\textwidth}p{0.09\textwidth}p{0.09\textwidth}p{0.09\textwidth}%
p{0.09\textwidth}}
\hline
Beam, $T_{\mbox{\tiny{kin}}}$ & Target & $K_{\mbox{\tiny{dc}}}$
 & \multicolumn{5}{c}{\rule{0mm}{5mm}$\frac{\Delta N(\Delta f)}{0.5 N_{\mbox{\tiny{tot}}}}$~(\%)} \\
(A$\cdot$GeV) &  & 
 & \multicolumn{5}{c}{for $\Delta f$ as $f_1$..$f_2$~(Hz)} \\
\hline
\multicolumn{3}{c}{}
 & \rule{0mm}{5mm}0..1 & $\Delta f_{\mbox{\tiny{ITS}}}$ & 49..51 & 99..101 & 149..151 \\
\hline
d, 0.25 & CH$_2$ & 0.56
 & 13.0 & 3.2 & 7.9 & 4.4 & 1.5 \\
d, 0.25 & C & 0.82
 & 19.5 & 1.1 & 1.0 & 0.26 & 0.08 \\
d, 3.5 & CH$_2$ & 0.926
 & 32.0 & 5.4 & 5.4 & 0.6 & 0.35 \\
d, 3.5 & C & 0.918
 & 30.1 & 0.9 & 5.9 & 1.1 & 0.53 \\
\hline
\end{tabular}
\end{table}

\begin{figure}[htb]

\vspace*{2mm}

\includegraphics[width=0.49\textwidth]{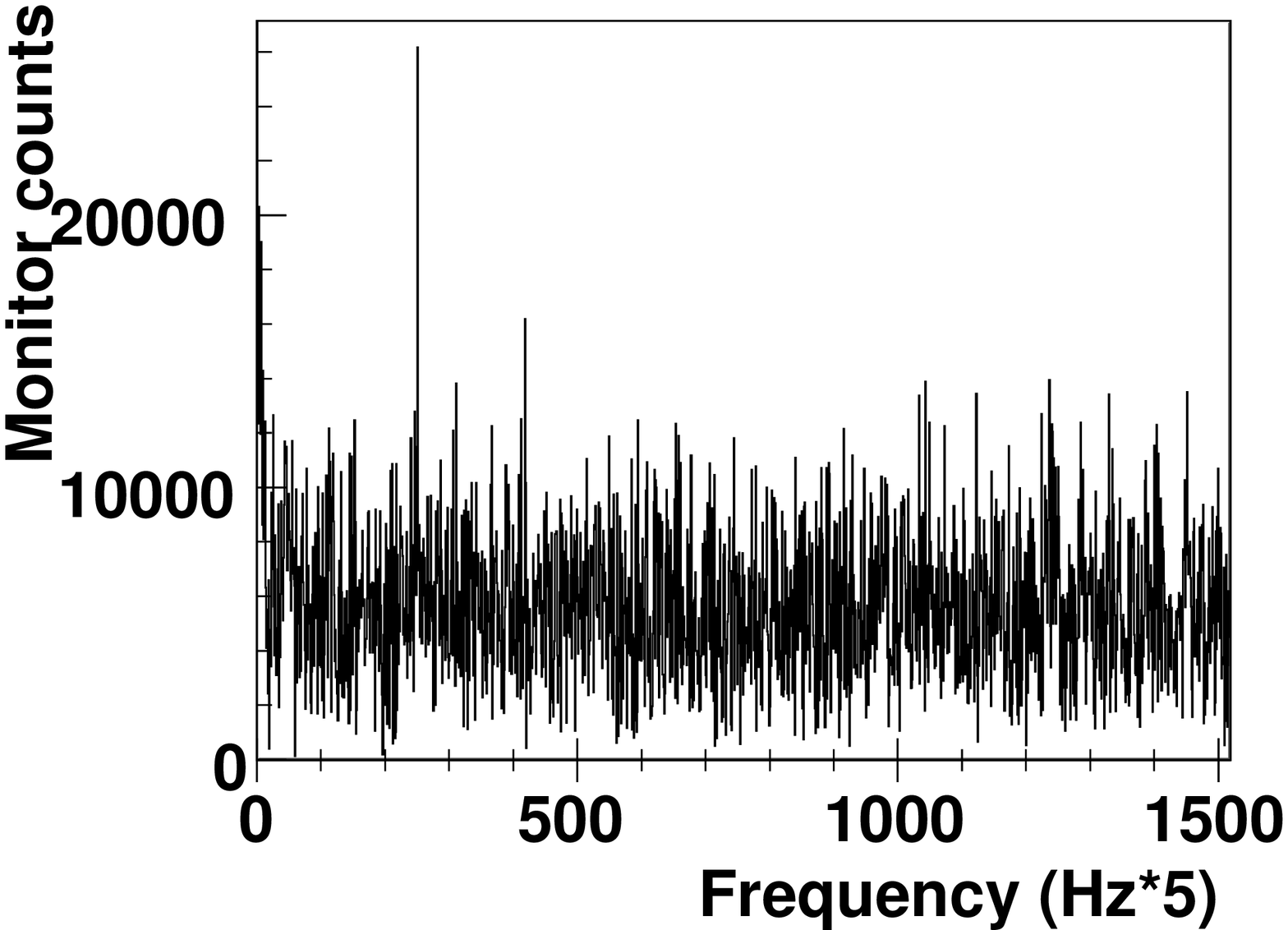}%
\includegraphics[width=0.49\textwidth]{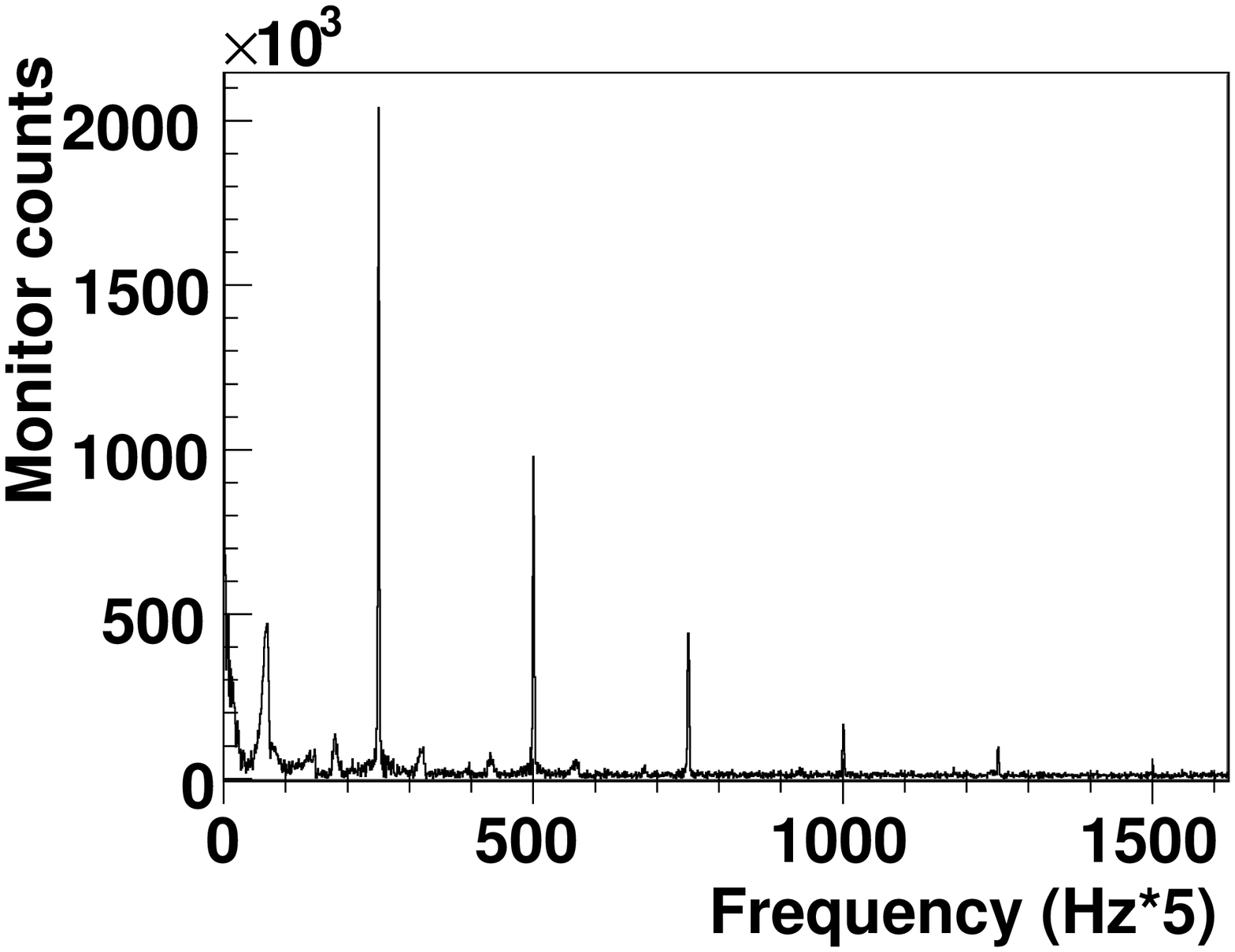}
\caption{The Fourier transform of the monitor counts
from the $^{106}$Ru radioactive source without (left) and with (right)
Nuclotron internal deuteron beam of
$T_{\mbox{\tiny{kin}}} = 150$~A$\cdot$MeV on the CH$_2$ target
for the whole run.
See text for description.}
\label{spill.fig.cnts_int150_betaCH2}
\end{figure}

\begin{figure}[htb]

\vspace*{2mm}

\includegraphics[width=0.49\textwidth]{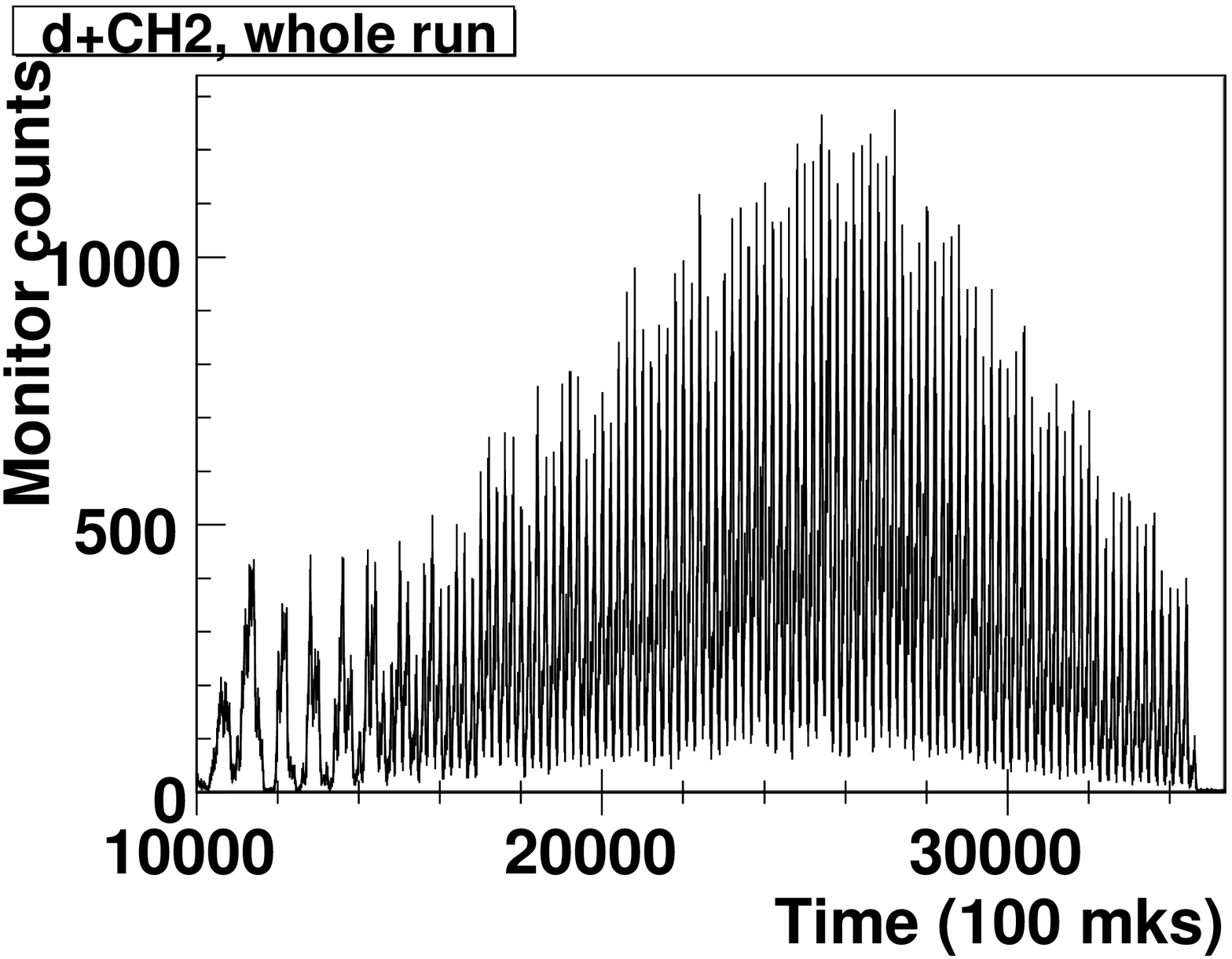}%
\includegraphics[width=0.49\textwidth]{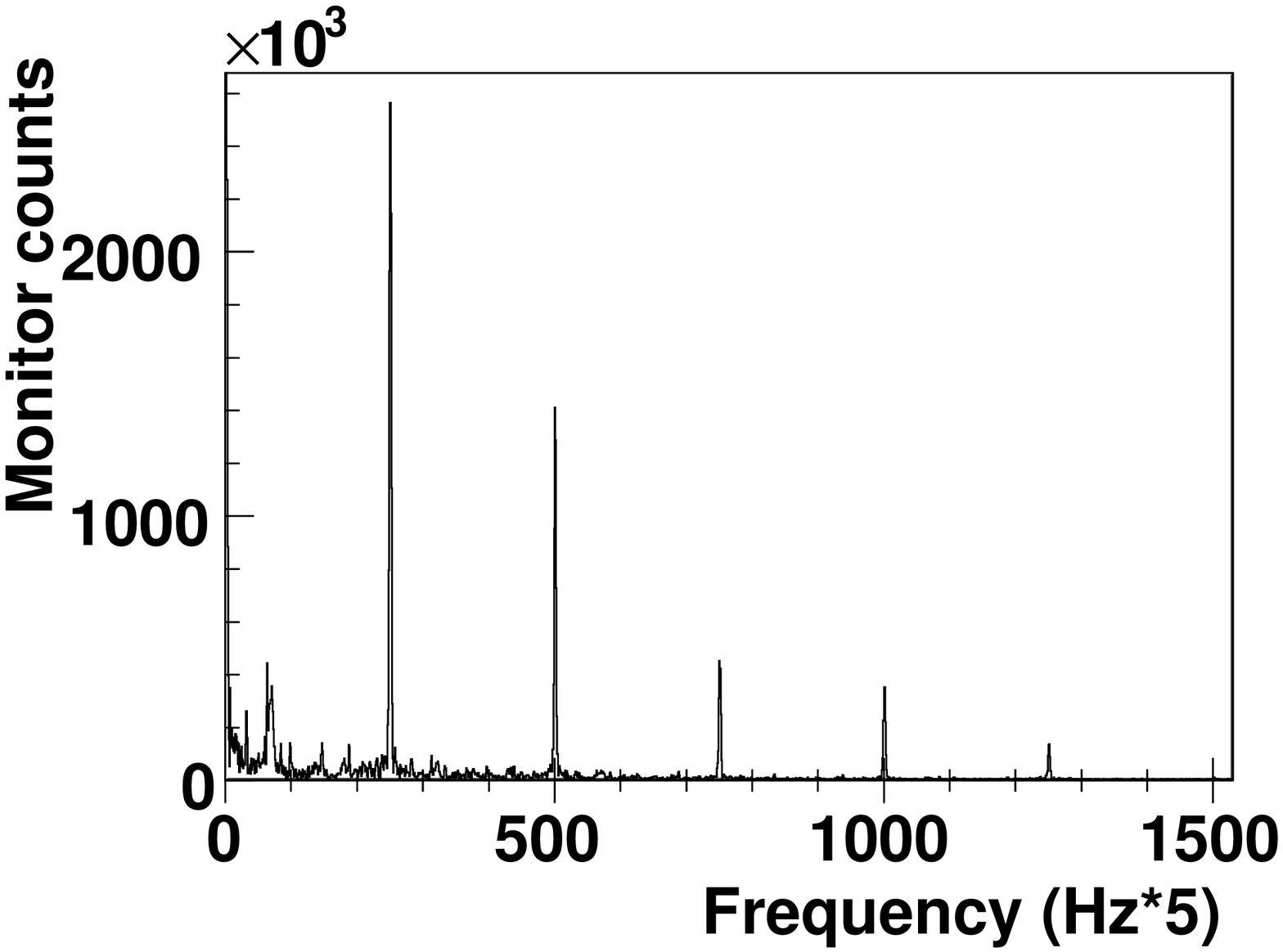}
\caption{The monitor counts during
the whole run (left) and their Fourier transform (right) for Nuclotron
internal deuteron beam of
$T_{\mbox{\tiny{kin}}} = 250$~A$\cdot$MeV on the CH$_2$ target.
See text for description.}
\label{spill.fig.cnts_int250_CH2}
\end{figure}

\begin{figure}[htb]

\vspace*{2mm}

\includegraphics[width=0.49\textwidth]{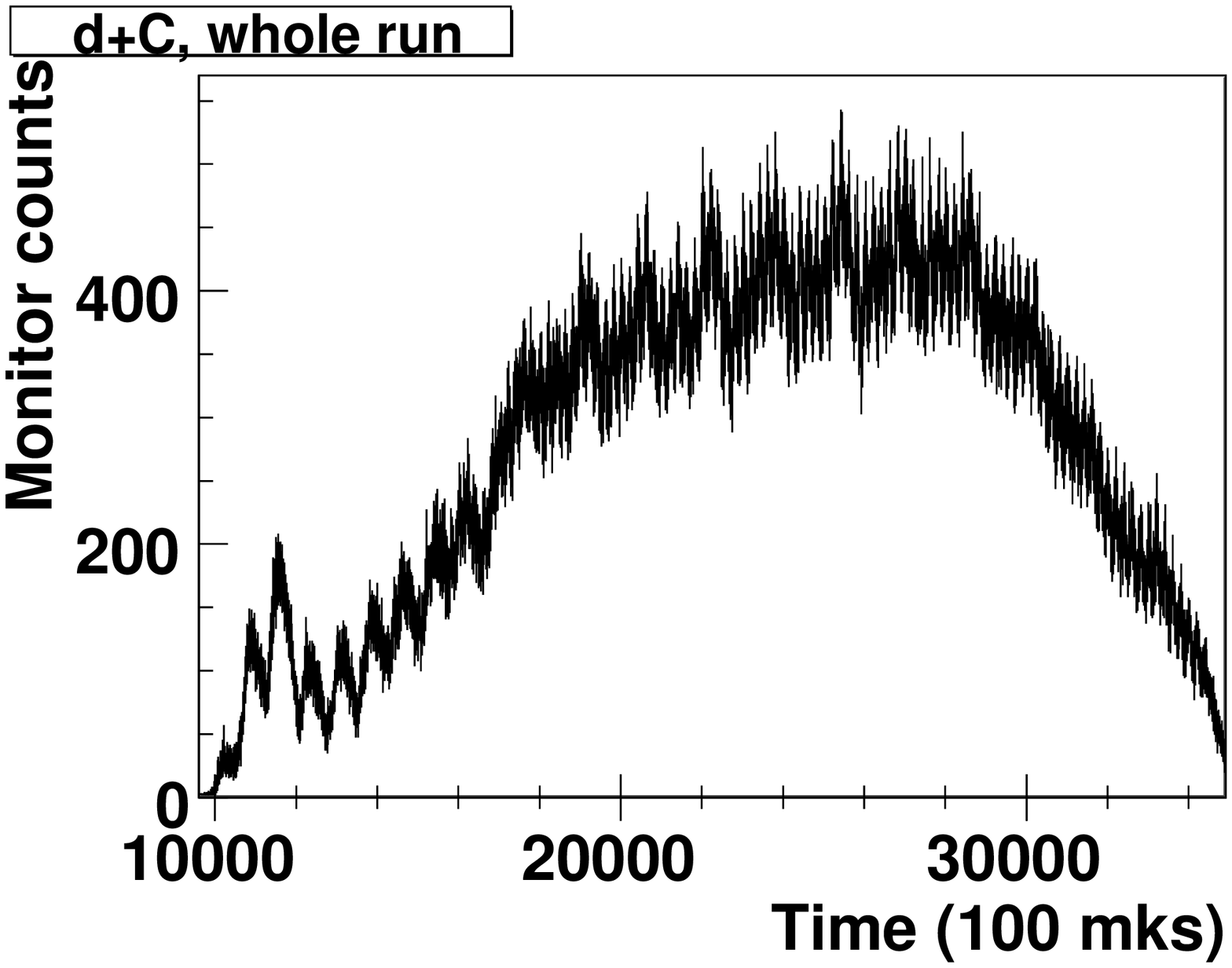}%
\includegraphics[width=0.49\textwidth]{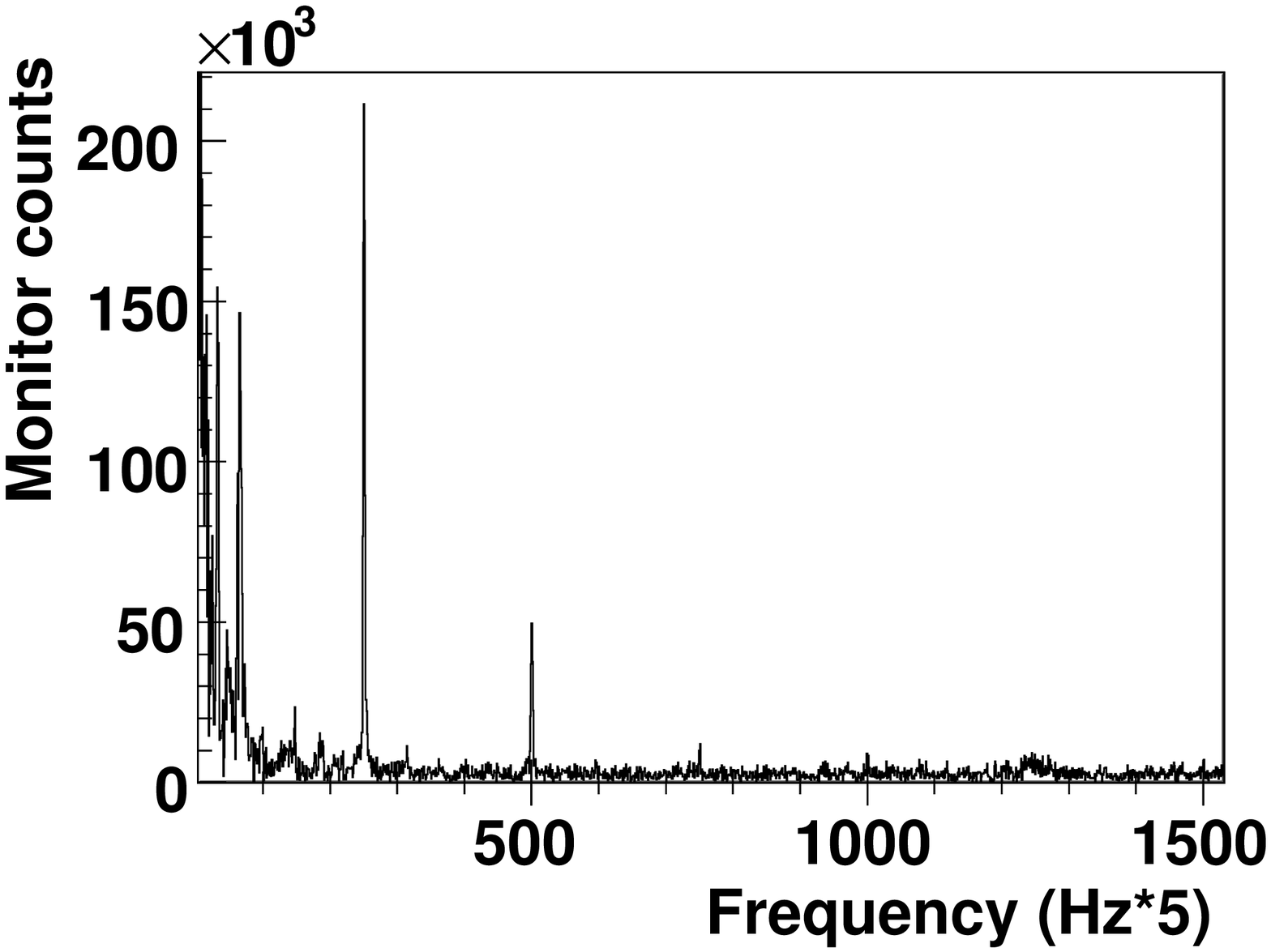}
\caption{The monitor counts during
the whole run (left) and their Fourier transform (right) for Nuclotron
internal deuteron beam of
$T_{\mbox{\tiny{kin}}} = 250$~A$\cdot$MeV on the C target.
See text for description.}
\label{spill.fig.cnts_int250_C}
\end{figure}

\begin{figure}[htb]

\vspace*{2mm}

\includegraphics[width=0.49\textwidth]{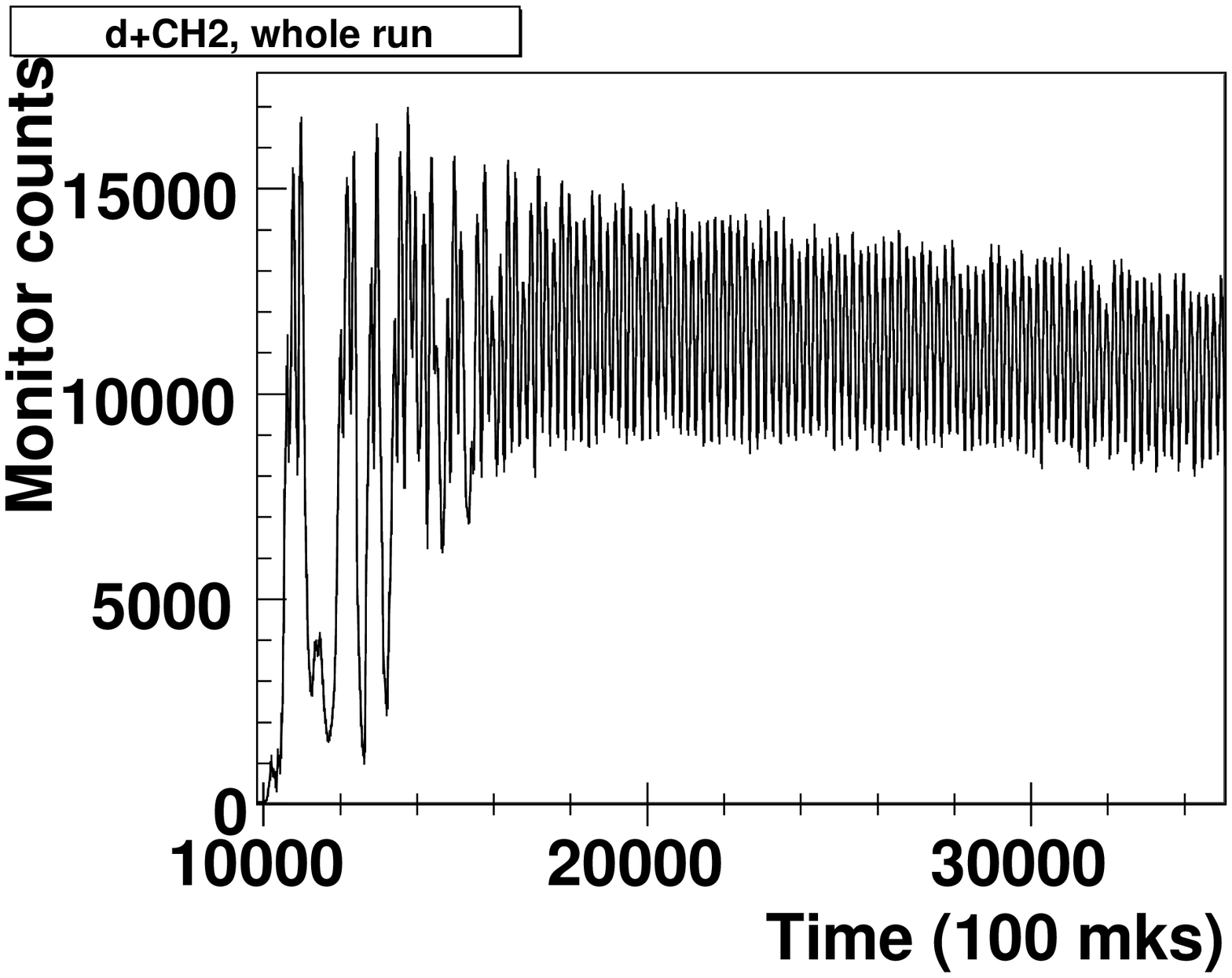}%
\includegraphics[width=0.49\textwidth]{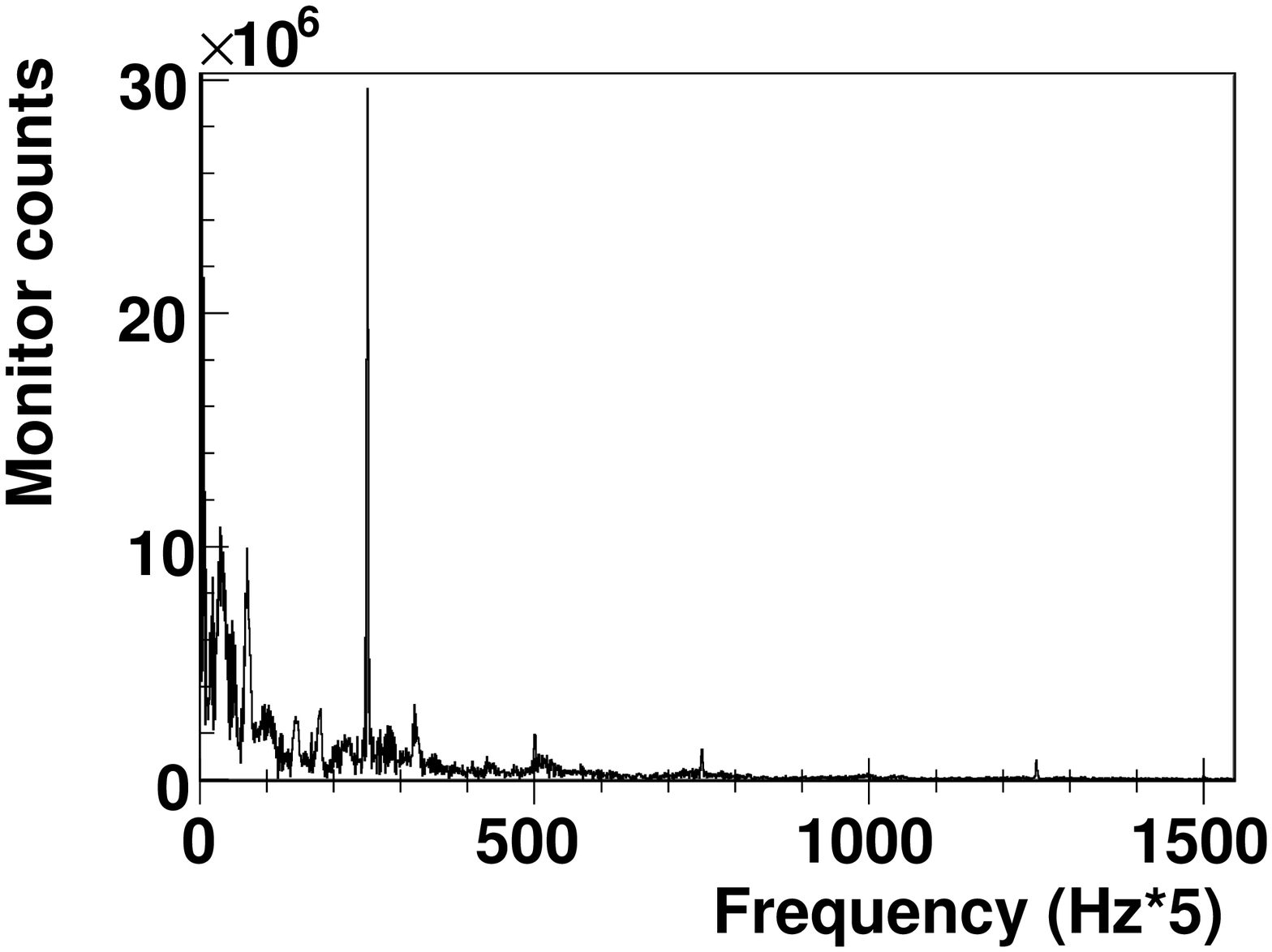}
\caption{The monitor counts during
the whole run (left) and their Fourier transform (right) for Nuclotron
internal deuteron beam of
$T_{\mbox{\tiny{kin}}} = 3.5$~A$\cdot$GeV on the CH$_2$ target.
See text for description.}
\label{spill.fig.cnts_int3500_CH2}
\end{figure}

\begin{figure}[htb]

\vspace*{2mm}

\includegraphics[width=0.49\textwidth]{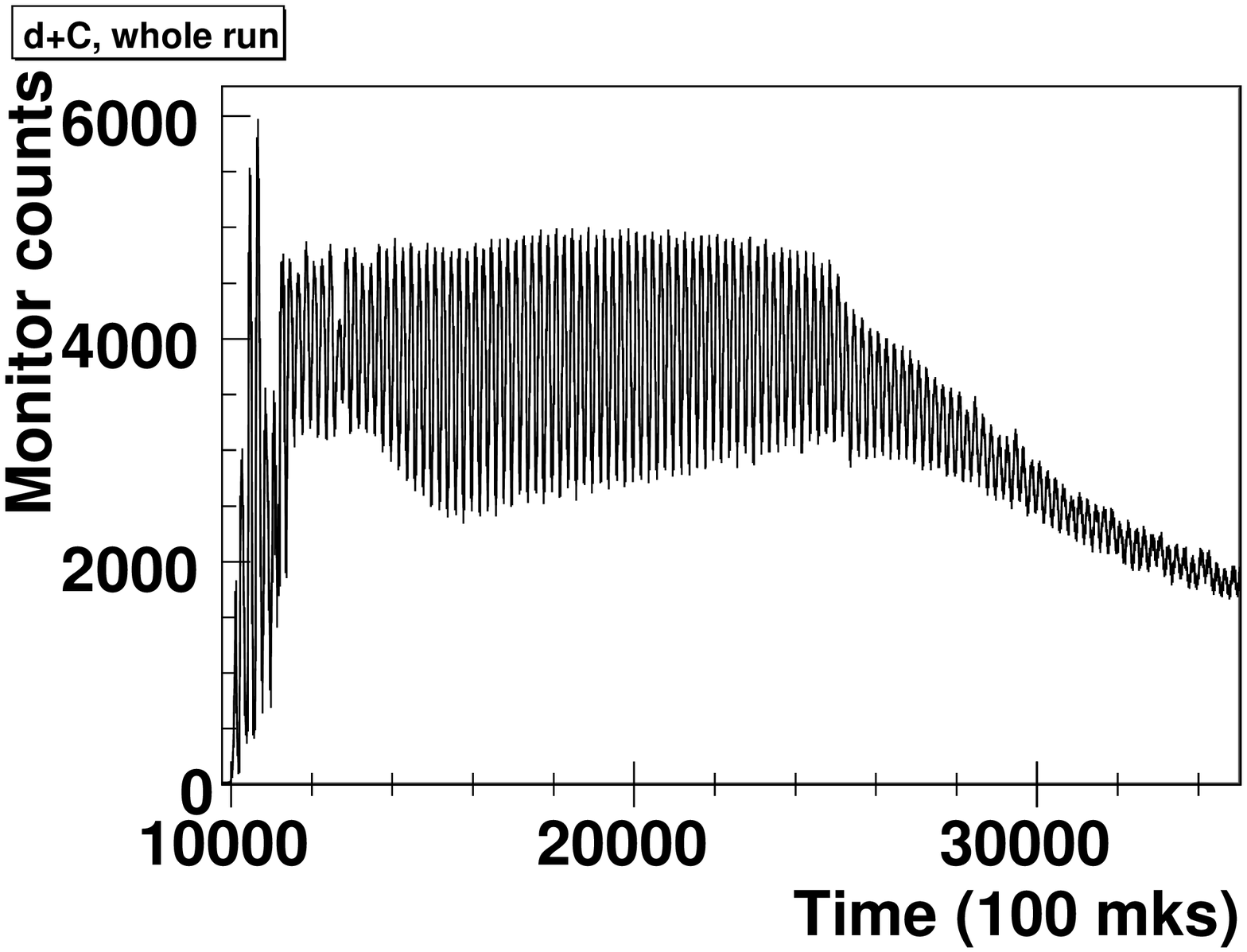}%
\includegraphics[width=0.49\textwidth]{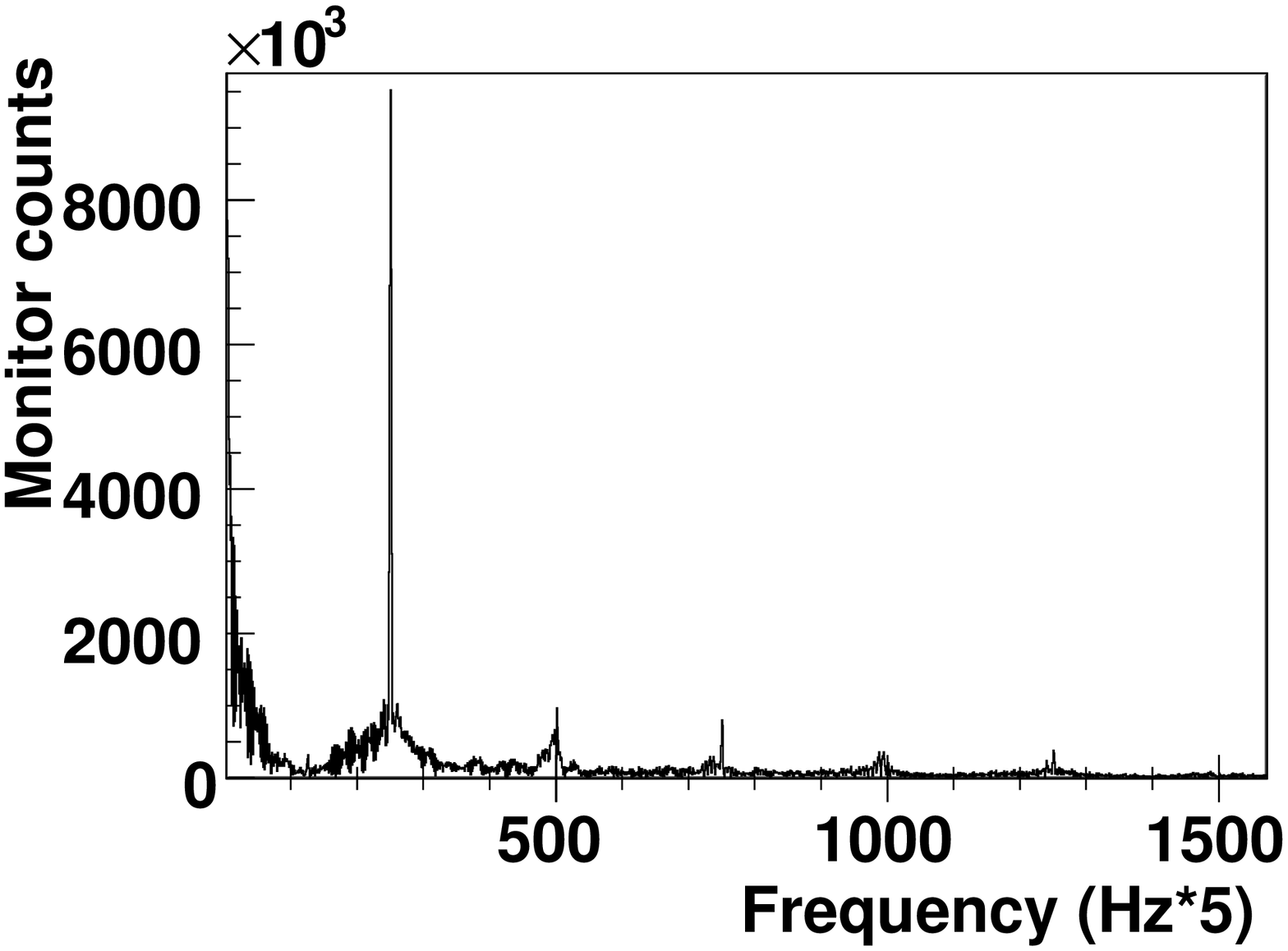}
\caption{The monitor counts during
the whole run (left) and their Fourier transform (right) for Nuclotron
internal deuteron beam of
$T_{\mbox{\tiny{kin}}} = 3.5$~A$\cdot$GeV on the C target.
See text for description.}
\label{spill.fig.cnts_int3500_C}
\end{figure}

\subsection{Spill DAQ system}
\label{spill.setup.sw}


Both setups have very similar software however each uses own hardware,
in particular the main host.
The overall Spill DAQ layout is pictured in
Fig.~\ref{spill.fig.schema}, where we can see the main host
under the FreeBSD OS as the rectangle entitled ``{\sffamily Spill DAQ}'',
and optional hosts --- as the rectangles
``{\sffamily visualization client}''.
This main host deals with the CAMAC hardware using the PK009 ISA adapter
for the KK009 CAMAC crate controller \cite{KK009}.
In the OS kernel context it has the
{\bfseries\itshape camac(4)} facility \cite{GrOllat},
{\bfseries\itshape spill(4)} CAMAC
kernel module,
as well as the modules in the
{\bfseries\itshape netgraph(4)} package \cite{netgraph4} style
(Fig.~\ref{spill.fig.graph}).
Fig.~\ref{spill.fig.graph} shows the data streams
transportation from the {\bfseries\itshape spill(4)} CAMAC kernel module
through the nodes, which are represented as
rectangles with node name (up), type (left), and ID (right). The data streams
are propagated along the graph edges, which are formed by pairs of
connected hooks \cite{netgraph4} (octagons in Fig.~\ref{spill.fig.graph}).
The data are
injected into graph by the \verb|camacsrc| node and demultiplexed by the
\verb|fifos| node into more than one identical streams, which could be transferred
both locally and remotely. In Fig.~\ref{spill.fig.schema} we can see
two of the data stream consumers: {\bfseries\itshape writer(1)}
and {\bfseries\itshape b2h(1)}, whose standard inputs are
fed by standard outputs of {\bfseries\itshape ngget(1)} instances, which
reads the \verb|fifos| local outputs through the
{\itshape netgraph} sockets provided by {\bfseries\itshape ng\_socket(4)} node
instances.
Each \verb|fifos| output as well as their input could be
connected and disconnected without disturbing other one(s).
Due to \verb|fifos| loading during the OS boot sequence all
\verb|load| user commands (see Table~\ref{spill.tbl.Make}) are independent
from each other.
The ``{\sffamily visualization client}'' in
Fig.~\ref{spill.fig.schema} executes one of the possible clients of the
{\bfseries\itshape b2h(1)} server --- the ROOT script
{\itshape client5.C}~.

The {\bfseries\itshape netgraph(4)} package \cite{netgraph4}
and {\itshape ngdp} framework
\cite{Isup_arXiv4474,Isup_arXiv4482}
involvement reduces the DAQ system
implementation efforts essentially,
because a number of software modules
are ready to use:
{\bfseries\itshape ng\_camacsrc(4)}, {\bfseries\itshape ng\_fifos(4)},
{\bfseries\itshape ngget(1)} \cite{Isup_arXiv4482},
{\bfseries\itshape writer(1)} \cite{IsupJINRC01-116},
{\bfseries\itshape ng\_ksocket(4)}, {\bfseries\itshape ng\_socket(4)}.
The produced data could be distributed
in form of both the {\itshape ngdp} packet stream(s) (for raw cycle-by-cycle
ones) and ROOT framework \cite{ROOTproc} \verb|TMessage|s (for
{\bfseries\itshape b2h(1)} produced
histograms). Specifically for the Spill DAQ we implement only
the {\bfseries\itshape spill(4)}, {\bfseries\itshape spillconf(1)},
{\bfseries\itshape spilloper(1)}, and {\bfseries\itshape b2h(1)}
software modules.

The {\bfseries\itshape spill(4)} module is intended to work with the
corresponding CAMAC hardware (see \ref{spill.setup.hw}).
It
complies with requirements
of the {\bfseries\itshape camac(4)} and {\bfseries\itshape ng\_ca\-mac\-src(4)},
so implements the CAMAC interrupt handler function. This handler
recognizes the following interrupt occurrences (events):\\
$\bullet$ begin of burst (BoB),\\
$\bullet$ odd and even trigger at end of 100~$\mu$s time slice.\\
In total two data packet types are produced
to contain as follows:
\verb|CYC_BEG| ---
the BoB timestamp,
\verb|CYC_END| --- the read-out scaler data as the sequence of 50000 16-bit
little-endian values to cover up to
5~s of burst. The \verb|CYC_END| packet type is produced at obtaining the
50000th end-of-slice interrupt.

The CAMAC hardware description and handling are separated from the
{\bfseries\itshape spill(4)} module's source and grouped together in the
single {\itshape spill\_hard\-wa\-re.h} header file.
In the current implementation it
uses macro interface {\bfseries\itshape kk(9)} specific for the KK009 crate
controller \cite{KK009} instead of the crate controller independent
one {\bfseries\itshape camac(9)}, because the former
interface allows us to
perform
each CAMAC cycle
slightly faster
 \cite{IsupJINRC03}.

The {\bfseries\itshape spill(4)} module can be configured at startup by the
{\bfseries\itshape spillconf(8)} utility and
controlled during operation by the {\bfseries\itshape spilloper(8)} one.
To operate with the Spill DAQ it has the configuration file
(named by default {\itshape \$NGDPHOME/etc/spillsv.conf}) in the
Makefile format (see also
{\bfseries\itshape make(1)}). This file establishes the correspondence between
the user commands
(``targets'' in {\bfseries\itshape make(1)} terminology)
and actions which should be performed. This file is textual and could be
revised easily. So typical user command look like the following:\\
\verb|make -f spillsv.conf RUNFILE=test loadw|\\
All the defined user commands we summarize in the Table~\ref{spill.tbl.Make}.

The {\bfseries\itshape b2h(1)} (for ``binary to histograms'') module is
intended for ROOT histograms filling from the packet (binary) data
representation
obtained from \verb|ng_fifos| node output. Due to simple nature of our setups
we use the compiled-in histograms configuration from {\itshape hconf.h} header
file. The 50000-bin histograms for at least last cycle and sum during current
run, and optionally their Fourier transforms are booked at
{\bfseries\itshape b2h(1)} startup. At each \verb|CYC_END| packet
arrival these histograms and some statistics are updated.
Namely, for 50 bins of 100~ms the mean of counts, its standard deviation,
dispersion and dispersion per mean are calculated and printed to standard
error output. The
{\bfseries\itshape b2h(1)} {\bfseries\itshape listen(2)}s on port 12342 for the
client registration. For each registered client the {\bfseries\itshape b2h(1)}
sends all histograms by \verb|TMessage|s once per cycle. Also all configured
histograms are written optionally to ROOT \verb|TFile| periodically or at
{\bfseries\itshape b2h(1)} termination as well as at \verb|SIGHUP| signal
obtaining. The \verb|SIGINT| signal resets all histograms, \verb|SIGUSR1|
and \verb|SIGUSR2| switches to read-and-discard
mode and vice versa. For user termination in
accuracy manner the \verb|SIGTERM| signal should be used.

The ROOT script {\itshape client5.C} connects to {\bfseries\itshape b2h(1)}
server, obtains \verb|TH1F| histograms encoded into \verb|TMessage|s, and
draws each of them in separate \verb|TCanvas|es. To allow the full
user interaction with these histograms the script should be terminated (i.e.,
by twice \verb|<Ctrl><C>| pressing), because the proper coexistence of two endless
loops (for \verb|TSocket| reading and X11 \cite{X11protocol} events handling)
could be managed only
in the executable like {\bfseries\itshape histGUI(1)}
 \cite{Isup_arXiv4482}.

\section{Nuclotron extracted beam results}
\label{spill.run_extr}

The
first setup version was implemented in 2009 and
successfully used
during 7 Nuclotron runs
in 2011--2015 to measure the extracted deuteron, carbon, and $^7$Li
beams time structure.
The scintillation counter ($100 \times 40 \times 4$~mm$^3$) with XP2020Q PMT
movable by motor was situated
in the beam halo
near the F4 focus of the VP1 transport beamline.
The PMT pulses shaped by 4F-115 discriminator were used as the
input monitor counts for the setup.

The measurements were performed during the data taking on the 
measurements of the soft photons yield \cite{Kokoulina2014}.
The typical averaged number of the counts in the 100~$\mu s$ time slice was 
28, 24 and 7 per beam spill for the $^7$Li, deuteron and carbon beam,
respectively.
In Figs.~\ref{spill.fig.cnts_ext_d}---\ref{spill.fig.cnts_ext_C}
the typical monitor counts proportional to the beam-target interaction rate
during a single cycle are shown
for the beam of
 deuterons,
$^7$Li, and
carbon with 3.5~A$\cdot$GeV kinetic energy. 
The abscissa is a time bin number, the ordinate is the number of counts
in the corresponding bin. The bin width is equal to the time slice,
100~$\mu$s. The typical 0.1~s time windows are shown on the right panels
instead of the full burst timescale.
The $K_{\mbox{\tiny{dc}}}$ value (see equation~(\ref{eq_Kdc})) could be obtained easily
from these data in
online or offline. For the single cycles depicted
in Figs.~\ref{spill.fig.cnts_ext_d}---\ref{spill.fig.cnts_ext_C}
the $K_{\mbox{\tiny{dc}}}$ during the full burst time $(t_2 - t_1)$ are
represented in Table~\ref{spill.tbl.kdc_ext}.
We are not show the whole run (per-bin cumulative sum) histograms because
the single cycle statistics is large enough.
The Fourier transforms of the corresponding
full burst statistics are shown in
Figs.~\ref{spill.fig.cnts_ext_d}---\ref{spill.fig.cnts_ext_C}
on the left panels.
As one can see, some frequencies (50~Hz and harmonics)
are emphasized. Their origin is a time structure of the slowly extracted beam.
The ratio $\frac{\Delta N(\Delta f)}{0.5 N_{\mbox{\tiny{tot}}}}$ values for
some set of frequency ranges $\Delta f = f_2 - f_1$ (in Hz) are represented
in Table~\ref{spill.tbl.kdc_ext}.
The $K_{\mbox{\tiny{dc}}}$ coefficient is smaller than 0.5. This reflects the
large width of the frequency spectrum (see right panels in
Figs.~\ref{spill.fig.cnts_ext_d}---\ref{spill.fig.cnts_ext_C}). On the other
hand, the contribution of the harmonics proportional to 50~Hz is lower than
7~\% in the worst case of $^7$Li. One can conclude, that significant
improvement of the extracted beam time structure is required for the high
intensity experiments, for instance, by the feedback from the monitor counter 
to the accelerator.

\section{Nuclotron internal beam results}
\label{spill.run_intr}

The
second setup version for the Nuclotron internal beam was implemented in 2014 and
successfully used during 2 Nuclotron runs (June'2014 and March'2015) to
measure the time scans of the beam-target interactions in the accelerator ring.
The scintillation counter ($100 \times 100 \times 4$~mm$^3$)
with FEU-30 PMT observes the beam-target interaction point of the Nuclotron
internal target station (ITS) \cite{IsupNIM13} from 1~m distance
under 60$^{\circ}$~angle in horizontal plane to produce
the input monitor counts for both ITS and current setup DAQ systems.
The analog signal from PMT with the rising time of $\sim 5$~ns and total 
duration of $\sim 15$~ns is shaped by the 4F-115 discriminator 
with $\sim 40$~mV threshold. The duration 
of the output pulse is $\sim 100$~ns.
In the second setup version
the counting
is started by hardware
at the same time offset for each cycle (with $\approx 50$~ns uncertainty)
 after the BoB event occurrence.
This allows us to work with
the whole run
histograms because the subtle counting
start variation in time can not ``smudge''
the time structure on such additive picture.

The quality of the electronic chain has been checked using $^{106}$Ru
radioactive (RA) source. The Fourier transform of the monitor counts
from the $^{106}$Ru RA source is shown on the left panel in
Fig.~\ref{spill.fig.cnts_int150_betaCH2} for the whole run.
These data were obtained while the 150~A$\cdot$MeV deuteron beam was accelerated
and circulated in the Nuclotron ring, however the internal target was not run.
The averaged number of counts in the 100~$\mu$s time slice is
$\sim 15$. The data demonstrate white noise picture that proofs the quality
of the used electronic chain. The right panel in 
Fig.~\ref{spill.fig.cnts_int150_betaCH2} shows the Fourier transform of the
monitor counts from both the deuterons of the same
energy on CH$_2$ target and the $^{106}$Ru RA source simultaneously.
The frequencies corresponding to target mechanical vibrations and to
the $n \cdot 50$~Hz harmonics are clearly seen.

The data on the deuteron beam time structure were obtained
simultaneously with the data taking for the DSS experiment
\cite{Ladygin:2014yla} with 10~$\mu$m CH$_2$ foil and 
8~$\mu$m carbon multiwire targets
with the typical intensity of $5 \div 8 \cdot 10^8$ deuterons per spill. The 
typical averaged number of the counts in the 100~$\mu$s time slice was 
$\sim 15$ and $\sim 180$ per spill at the beam energy of 250~A$\cdot$MeV
and 3.5~A$\cdot$GeV, respectively.

In Figs.~\ref{spill.fig.cnts_int250_CH2} and \ref{spill.fig.cnts_int250_C}
we can see the full burst timescale
 (left panels) of the monitor counts
during
the whole run (per-bin cumulative sum) from the beam of deuterons with
250~A$\cdot$MeV kinetic energy interacting with the
CH$_2$ and
C targets, respectively.
The Fourier transforms of the corresponding
full burst
statistics for the CH$_2$
and C target are shown in Figs.~\ref{spill.fig.cnts_int250_CH2}
and \ref{spill.fig.cnts_int250_C} on right panels.
The same for the 3.5~A$\cdot$GeV deuteron beam is shown in
Figs.~\ref{spill.fig.cnts_int3500_CH2} and \ref{spill.fig.cnts_int3500_C}.
One can see some frequencies (50~Hz and harmonics) other than originated
from the mechanical target vibrations
are enhanced, however their relative intensity is decreased with the frequency
increasing. Their origin can be a space-time structure of the Nuclotron internal
beam.
The ratio $\frac{\Delta N(\Delta f)}{0.5 N_{\mbox{\tiny{tot}}}}$ values for
some set of enhanced frequencies $f$ in narrow ranges $\Delta f = f \pm 1$~Hz
are represented in Table~\ref{spill.tbl.kdc_int}.
The ``$\Delta f_{\mbox{\tiny{ITS}}}$'' column contains this ratio for
mechanical vibrations
frequency of the internal target ($\approx 13$~Hz for C and $\approx 14$~Hz
for CH$_2$ in the represented data).
The results confirm the earlier observation \cite{IsupNIM13} that the
mechanical vibrations are larger for CH$_2$ target. The $K_{\mbox{\tiny{dc}}}$
coefficient is $\sim 0.9$ at 3.5~A$\cdot$GeV for both targets. The
contribution of the $n \cdot 50$~Hz harmonics is only about $8 \div 10$~\%.
However we observe the significant difference in $K_{\mbox{\tiny{dc}}}$
coefficient and contribution of the $n \cdot 50$~Hz harmonics for CH$_2$
and C targets at low energies. For carbon target the
$K_{\mbox{\tiny{dc}}} \sim 0.8$ and
the $n \cdot 50$~Hz harmonics contribution is only 1.5~\%. Therefore, one can
conclude that the time structure of the internal beam interaction with
carbon target
allows one to perform experiments at ITS. For CH$_2$ target the
$K_{\mbox{\tiny{dc}}}$ is only 0.56 and the $n \cdot 50$~Hz harmonics
contribution is $\sim 15$~\%. The reason of this difference is the subject of
further studies.

\section*{Conclusions}
\label{spill.concl}

The results of the present work can be summarized as follows:\\
$\bullet$ Two similar setups for precise measurements of the Nuclotron beam time
structure are designed using standard CAMAC modules produced at LHEP and DLNP.
They are commissioned for the experiments at internal target and at the slowly extracted beam. 
Both setups are working with the 
100~$\mu$s time slices, so they are able to investigate 
the frequency domain up to $\approx 4.5$~kHz.\\  
$\bullet$
The data on the beam time structure could be easily distributed online
through network on the cycle-by-cycle basis 
due to Spill DAQ design on base of the {\itshape ngdp} framework
\cite{Isup_arXiv4474,Isup_arXiv4482}. 
Therefore, these setups are suitable to work continuously
during the whole Nuclotron run
and provide data for the beam quality estimations.\\
$\bullet$
The first setup version was successfully used to measure the time 
structure of the slowly extracted beam of deuterons, $^7$Li, and carbon at 
3.5~A$\cdot$GeV. \\
$\bullet$ The second setup version was used to measure the time structure of the
beam-target interaction inside the Nuclotron ring at ITS \cite{oldIntTarg,IsupNIM13}.
It has been found the influence of the mechanical vibrations of the internal target
on the data taking. \\
$\bullet$ The developed setups can be efficiently used as a tool for the
permanent monitoring of the Nuclotron beam quality during the experiments
performed at Nuclotron.

\section*{Acknowledgments}

The authors have the pleasure to thank Prof. L.S.Zolin for initiation of
this work, cooperation during measurements, and permanent interest to results,
the Nuclotron team and in particular Dr. A.V.Butenko --- for fruitful
cooperation. The authors are grateful to 
A.N.~Livanov, S.M.~Piyadin, and A.N.~Khrenov for the technical support 
during the measurements at internal target.  
The present work was supported in part by the RFBR Grant no.13-02-00101a.


\end{document}